\documentclass[aps,twocolumn,showpacs]{revtex4}
\usepackage{graphicx}

\begin{document}

\title{Magnetization of potassium doped p-terphenyl and p-quaterphenyl by high pressure synthesis}

\author{Wenhao Liu, Hai Lin, Ruizhe Kang, Yue Zhang, Xiyu Zhu$^*$, and Hai-Hu Wen}\email{xyzhu@nju.edu.cn, hhwen@nju.edu.cn}

\affiliation{Center for Superconducting Physics and Materials,
National Laboratory of Solid State Microstructures and Department
of Physics, National Center of Microstructures and Quantum
Manipulation, Nanjing University, Nanjing 210093, China}

\date{}

\begin{abstract}
By using high pressure synthesis method, we have fabricated the potassium doped para-terphenyl. The temperature dependence of magnetization measured in both zero-field-cooled and field-cooled processes shows step like transitions at about 125 K. This confirms earlier report about the possible superconductivity like transition in the same system. However, the magnetization hysteresis loop exhibits a weak ferromagnetic background. After removing this ferromagnetic background, a Meissner effect like magnetic shielding can be found. A simple estimate on the diamagnetization of this step tells that the diamagnetic volume is only about 0.0427\% at low temperatures, if we assume the penetration depth is much smaller than the size of possible superconducting grains. This magnetization transition does not shift with magnetic field but is suppressed and becomes almost invisible above 1.0 T. The resistivity measurements are failed because of an extremely large resistance. By using the same method, we also fabricated the potassium doped para-quaterphenyl. A similar step like transition at about 125 K was also observed by magnetization measurement. Since there is an unknown positive background and the diamagnetic volume is too small, it is insufficient to conclude that this step is derived from superconductivity although it looks like.
\end{abstract}

\pacs{74.25.-q, 74.70.-b, 74.70.Kn} \maketitle

\section{Introduction}

Organic superconductors have been extensively studied, due to the presence of possible high temperature superconductivity and exotic superconducting pairing mechanism. Usually, the organic superconductors are accompanied by low-dimensional structures, rich magnetic phase transitions and novel ground states. In most cases, the band width of organic superconductors is supposed to be very narrow, which inevitably leads to the strong correlation effect. The early discovered organic superconductors mainly include organic salts\cite{Ogs1,Ogs2}. Lately, intercalated graphite compounds, metal doped fullerenes, and aromatic compounds are found superconductive. Therefore, in general, there are mainly two kinds of organic superconductors: organic salts such as (TMTSF)$_2$X\cite{Ogs1} and (BEDT-TTF)$_2$X\cite{Ogs2} and metal intercalated organic materials\cite{OGSCReview}. For quasi one dimensional (TMTSF)$_2$X, as the first discovered organic superconductor, superconductivity emerges at 0.9 K under 12 kbar\cite{Ogs1}. Yet, the highest superconducting T$_c$ of two dimensional (BEDT-TTF)$_2$X can rise up to 14.2 K\cite{13KOgs2}. In this system, anti-ferromagnetism, charge density wave, Mott insulator state and superconductivity can be tuned by adjusting pressure and temperature. Graphite intercalation compounds including KC$_8$\cite{KC8} and CaC$_6$\cite{CaC6} show superconducting transitions below 10 K. While, In the potassium doped fullerenes\cite{furllerenes}, with the highest T$_c$ up to 40 K, some characteristics of strong correlated electronic systems were found, like metal-insulating transition, antiferromagnetic spin fluctuation etc.\cite{MIT,SPFlu}

Recently, the superconductors of alkali metal doped aromatic compounds attract great attentions as a new organic superconducting system. Since K$_{3.3}$Picene with T$_c$=18 K was discovered in 2010\cite{K3picene}, several new superconductors have been found, such as K$_x$Phenanthrene with T$_c$ of 5 K\cite{Kxphe}, K doped 1,2:8,9-dibenzopentacene with the claim of the highest T$_c$ of 33 K\cite{Kx1289}. Very recently, it was reported that high temperature superconductivity with T$_c$ above 120 K might exist in potassium doped \emph{p}-terphenyl\cite{chenxiaojia120K}. A superconducting gap like feature was then observed in the potassium dosed single crystal of \emph{p}-terphenyl by measurements of angle resolved photoemission spectroscopy (ARPES)\cite{Dessau}. This gapped feature gradually disappears with increasing of temperature. The gapped feature was also observed in the potassium dosed monolayer film of \emph{p}-terphenyl\cite{FengDL}, however the authors find that the gapped feature does not change with magnetic field, suggesting a non-superconductive origin of this gap. In the original paper that reports the possible superconductivity, the magnetic shielding volume fraction estimated from the step like magnetization is quite small and no resistivity data was reported. As far as we know, no second group reports the observation of the phenomenon. In order to confirm the superconductivity in this system, we have fabricated the potassium doped \emph{p}-terphenyl under both ambient pressure and high pressure (2 GPa). For the samples synthesized under ambient pressure, we have not observed the step like transitions of magnetization around 120 K. However, for high pressure synthesized samples, we can easily reproduce the similar feature. Surprisingly, on the potassium doped \emph{p}-quaterphenyl we also observed this step like magnetic transition at about 125 K. In this paper, we report the synthesis and magnetization measurements of these samples.

\section{Experimental details}
Potassium metal was cut into pieces and mixed with \emph{p}-terphenyl or \emph{p}-quaterphenyl in a molar ratio 3:1 in the glove box filled with argon (O$_2$ less than 0.1ppm, H$_2$O less than 0.1ppm). The mixture was then put into Al$_2$O$_3$ crucible and sealed in quartz tube under high vacuum. The quartz tubes were heated up to 523 K for \emph{p}-terphenyl, and 623 K for \emph{p}-quaterphenyl, respectively. We kept sintering the samples with these two temperatures for 3 days. Then, we obtained the ambient pressure synthesized samples. In order to improve the quality and compactness of samples, we also synthesized the samples with high pressure. For high pressure synthesis, the raw materials were pressed into pellet of 4 mm in diameter and 7 mm of length. Each pellet was placed into a BN container, surrounded by a graphite sleeve resistance heater and pressure transmitting MgO rods. Afterwards, the assemblies were submitted to 2 GPa at room temperature. Subsequently, they were heated up to 523 K for \emph{p}-terphenyl, and 623 K for \emph{p}-quaterphenyl, respectively. The temperatures were kept for 10 hours under these pressures, before quenched to room temperature. Finally, we released the pressure. The final products after high pressure synthesis are black and sensitive to air. The magnetization results were measured by a Quantum Design instrument SQUID-VSM with magnetic field up to 7 T, and the magnetic field was applied parallel to the axis of pellets in the measurements.

\section{Results and discussion}
\begin{figure}
\includegraphics[width=8.5cm]{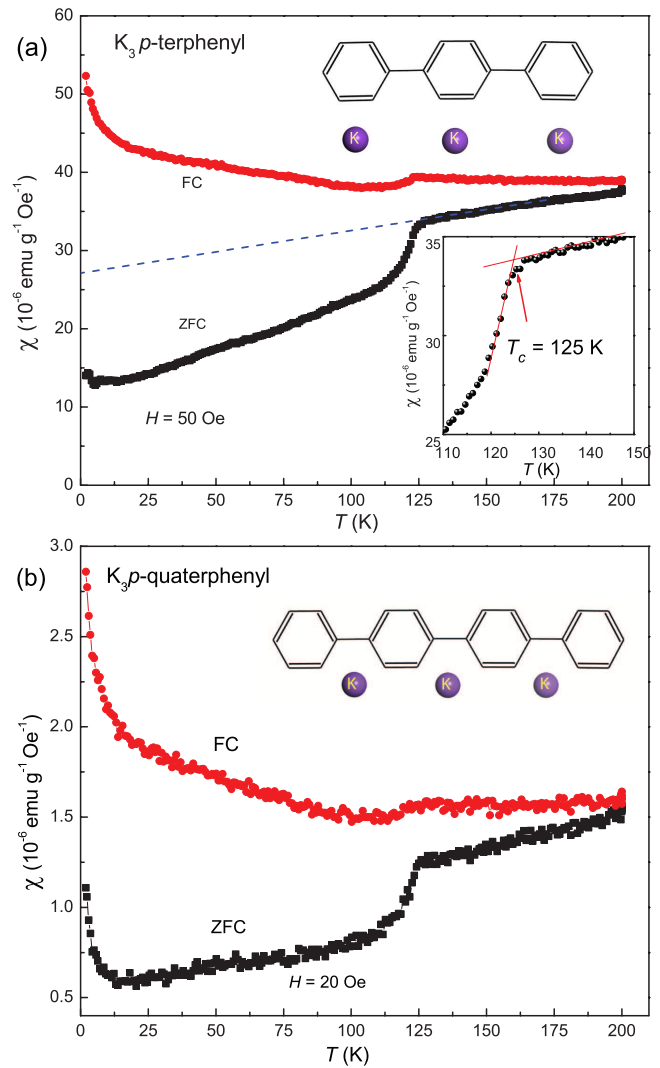}
\caption{(Color online) Temperature dependence of magnetic susceptibility measured in both ZFC and FC modes for K$_3$\emph{p}-terphenyl under 50 Oe and K$_3$\emph{p}-quaterphenyl under 20 Oe, respectively. Upper rights present the schematic molecular structure of potassium doped \emph{p}-terphenyl and \emph{p}-quaterphenyl. The inset of Fig.1(a) shows the enlarged view of magnetic susceptibility for K$_3$\emph{p}-terphenyl near the step transition temperature.} \label{fig1}
\end{figure}
In Fig.~\ref{fig1}, we present the temperature dependence of magnetic susceptibility measured in the zero-field-cooled (ZFC) and field-cooled (FC) modes for K$_3$\emph{p}-terphenyl under 50 Oe and K$_3$\emph{p}-quaterphenyl under 20 Oe, respectively. Step like transitions can be clearly seen near 125 K in the magnetic measurements for both compounds, which confirms earlier report about the possible superconductivity like transition in K$_3$\emph{p}-terphenyl\cite{chenxiaojia120K}. However, below the temperature of the step like transition, our data of magnetic susceptibility is positive and exhibit a strong upturn at low temperatures. In order to compare with the shielding fraction with the previous work, $\Delta\chi$ has been determined as the difference of the zero-field-cooled data deducted with the extension line of normal state, which is about 1.36$\times$10$^{-5}$ emu g$^{-1}$ Oe$^{-1}$ at 4 K. This is about 22 times larger than previous report under the same magnetic field\cite{chenxiaojia120K}. Supposing a density of about 2.5 g/cm$^3$, we have 4$\pi\Delta\chi$ = -0.0427\%. As shown in Fig. 1(b), a similar behavior of magnetization has also been observed in K$_3$\emph{p}-quaterphenyl under 20 Oe, which possibly indicates the same origin for the transition in this compound. Upper rights of both figures show the schematic molecular structure of potassium doped \emph{p}-terphenyl and \emph{p}-quaterphenyl. But the real structures of the molecular crystals are still unknown. In inset of Fig. 1(a), an enlarged view of magnetic susceptibility for K$_3$\emph{p}-terphenyl near the step transition temperature is shown, which clearly presents the transition at about 125 K.

\begin{figure}
\includegraphics[width=8.5cm]{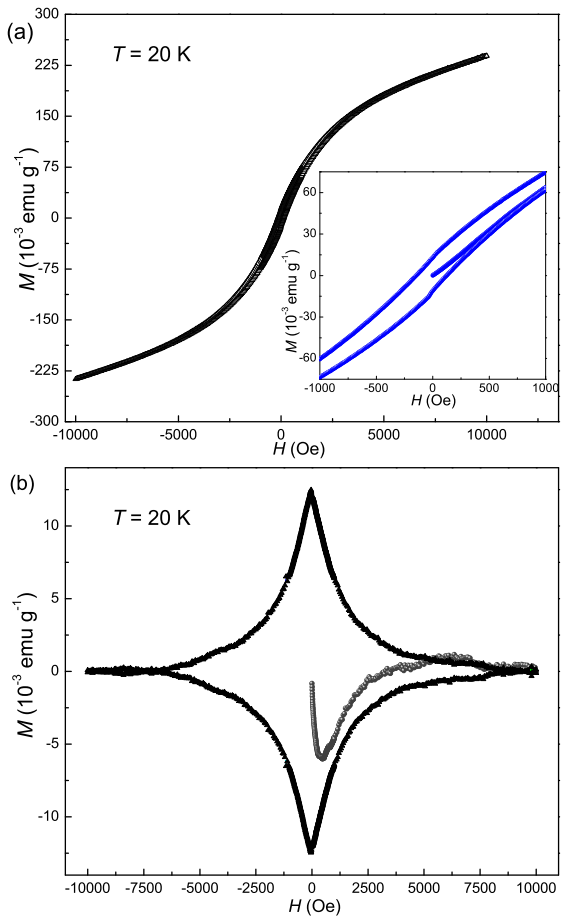}
\caption{ (Color online) (a) Magnetization hysteresis loop (MHL) of K$_3$\emph{p}-terphenyl at 20 K. The inset shows the enlarged view of MHL from -1000 Oe to 1000 Oe (b) MHL after deducting the normal state weak ferromagnetic background.
} \label{fig2}
\end{figure}

Fig.~\ref{fig2}(a) presents the magnetization hysteresis loop (MHL) of K$_3$\emph{p}-terphenyl up to 1 T at 20 K. As we can see in the inset of Fig. 2(a), the MHL shows a weak ferromagnetic signal together with a small hysteresis. This weak ferromagnetic signal may be induced by some impurities, or it is an intrinsic feature of the system. We take the averaged value between the increasing and decreasing fields as the background, and after removing the background, a Meissner effect like magnetic shielding has been observed, as shown in Fig. 2(b). This Meissner effect like magnetic shielding may reveal that it is a type-II superconductor. The phenomenon is very similar to the superconducting (Li${_{1-x}}$Fe${_x}$)OHFeSe crystal, which also has a weak ferromagnetic back ground\cite{LiOHFeSe} due to probably the Fe atoms at the Li sites. As we already estimated that the maximum diamagnetic signal at 4 K of K$_3$\emph{p}-terphenyl is about 4$\pi\chi$ $\approx$ -0.0427\%. This calculation is based on the assumption that the London penetration depth is much smaller than the size of the possible superconducting grains. If the real situation is not like that, namely the penetration depth is comparable or larger than the possible superconducting grain size, the calculated volume fraction will be modified to larger values. This indicates there might be some superconducting like phase in the sample, however the volume fraction is quiet small.

\begin{figure}
\includegraphics[width=8.5cm]{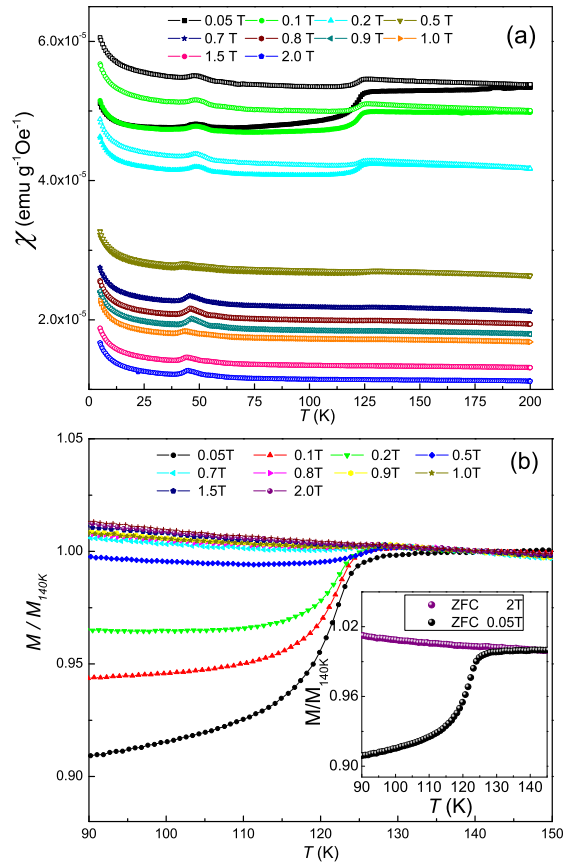}
\caption{(Color online) (a) Temperature dependence of magnetic susceptibility measured in both ZFC(solid points) and FC(hollow points) modes for K$_3$\emph{p}-terphenyl under different magnetic fields. (b) Temperature dependence of magnetic susceptibility divided by their own values at 140 K in ZFC mode. The inset shows the enlarged view of magnetic susceptibility under 0.05 T and 2 T near the phase transition temperature. } \label{fig3}
\end{figure}
In Fig.~\ref{fig3}, we present the temperature dependence of magnetic susceptibilities measured in ZFC and FC modes for K$_3$\emph{p}-terphenyl under different external fields. As shown in Fig. 3(a), the susceptibility presents the step like transition clearly near 125 K in the low magnetic fields. The peak like feature at about 47 K is due to the frozen oxygen in the sample or sample holder, which is a common problem of the SQUID measurement if some samples contain or absorb oxygen. The step of the transition is suppressed gradually with increasing magnetic field, and finally disappears above about 1.0 T. However, the transition temperature keeps almost unchanged. This is consistent with the previous report\cite{chenxiaojia120K}.  One possibility is that the magnetic field is insufficient to break Cooper pairs, but suppresses the Josephson coupling or the phase coherence among the possible superconducting grains. In order to present the transition more clearly, in Fig. 3(b), we show the temperature dependence of the normalized magnetic susceptibilities in ZFC mode from 90K to 150K. The normalization is done by dividing the magnetic susceptibilities by their own value at 140 K. As we can see, below 1.0 T, the magnetic field suppresses the transition step gradually, while the transition temperature is almost unchanged. Above 1.0 T, the step transition disappears totally. The inset of Fig. 3(b) shows the normalized magnetic susceptibilities under 0.05 T and 2 T. As shown, no phase transition can be observed under 2 T.

We must emphasize that, the step like transition of magnetic susceptibility at about 125 K and related analysis suggest that it is consistent with the picture of possible superconductivity. However since there is a positive background and the magnitude of the step like transition is still very weak, it is insufficient to conclude that the step is definitely derived from superconductivity. We also tried to measure resistivity but failed due to the very large resistance. It requires more time and efforts to prove that the phenomenon arises from superconductivity.

\section{Conclusions}

In summary, we have fabricated potassium doped \emph{p}- terphenyl and \emph{p}-quaterphenyl by high pressure method. The magnetic susceptibilities show the step like transitions near 125 K for both compounds, which confirms earlier report about the possible superconductivity like transition in potassium doped \emph{p}-terphenyl. After removing the week ferromagnetic background, a type-II superconductivity like magnetization hysteresis loop has been observed. With increasing the external field, the step is suppressed gradually. And above 1 T, the transition is completely smeared out. Based on the magnetic measurements, we conclude that there might be some small fraction of superconducting phase in potassium doped \emph{p}-terphenyl and \emph{p}-quaterphenyl. However, since there is a weak ferromagnetic background and the magnetic shielding volume fraction is too small, we would not conclude that this phenomenon is derived from superconductivity at this stage.

\section*{ACKNOWLEDGMENTS}
This work was supported by the Ministry of Science and Technology of China (Grant No. 2016YFA0300404, 2016YFA0401700, 2012CB821403), and the National
Natural Science Foundation China (NSFC) with the projects: A0402/11534005, A0402/11190023, A040206/11674164.


\begin{thebibliography}{00}

\bibitem{Ogs1} D. J\'erome, A. Mazaud, M. Ribault, and K. Bechgaard, J. Phys. Lett. (Pairs) \textbf{41}, 95 (1980).
\bibitem{Ogs2} A. M. Kini, U. Geiser, H. H. Wang, K. D. Carlson, J. M. Williams, W. K. Kwor, K. G. Vandervoort. J. E. Thompson, and D. L. Stupka, Inorg. Chem. \textbf{29}, 2555 (1990).
\bibitem{OGSCReview} Y. Kubozono, H. Mitamura, X. Lee, X. He, Y. Yamanari, Y. Takahashi, Y. Suzuki, Y. Kaji, R. Eguchi, K. Akaike, T. Kambe, H. Okamoto, A. Fujiwara, T. Kato, T. Kosugi and H. Aoki, Phys. Chem. Chem. Phys. \textbf{13}, 16476 (2011)
\bibitem{13KOgs2} H. Taniguchi, M. Miyashita, K. Uchiyama, K. Satoh, N. M\^ori, H. Okamoto, K, Miyagawa. K. Kanoda, M. Hedo, Y. Uwatoko, J. Phys. Soc. Jpn. \textbf{72}, 468 (2003).
\bibitem{KC8} N. B. Hannay, T. H. Geballe, B. T. Matthias, K. Andres, P. Schmidt, and D. MacNair, Phys. Rev. Lett. \textbf{14}, 225 (1965).
\bibitem{CaC6} N. Emery, C. H\'erold, M. d$^,$Astuto, V. Garcia, C. Bellin, J. F. Mar\^ech\'e, P. Lagrange, and G. Loupias, Phys. Rev. Lett. \textbf{95}, 087003 (2005).
\bibitem{furllerenes} A. Y. Ganin, Y. Takabayashi, Y. Z. Khimyak, S. Margadonna, A. Tamai, M. J. Rosseinsky, and K. Prassides, Nat. Mater. \textbf{7}, 367,(2008).
\bibitem{MIT} O. Gunnarsson, Rev. Mod. Phys. \textbf{69}, 575 (1997).
\bibitem{SPFlu} A. Y. Ganin, Y Takabayashi, P Jegli\^c, D. Ar\^con,	A. Poto\^cnik, P. J. Baker, Y. Ohishi, M. T. McDonald, M. D. Tzirakis, A. McLennan, G. R. Darling, M. Takata,	M. J. Rosseinsky and K. Prassides, Nature (London) \textbf{466}, 221 (2010).
\bibitem{K3picene} R. Mitsuhashi, Y. Suzuki, Y. Yamanari, H. Mitamura, T. Kambe, N. Ikeda, H. Okamoto, A. Fujiwara, M. Yamaji, N. Kawasaki, Y. Maniwa and Y. Kubozono, Nature (London) \textbf{464},76 (2010).
\bibitem{Kxphe} X. F. Wang, R. H. Liu, Z. Gui, Y. L. Xie, Y. J. Yan, J. J. Ying, X. G. Luo, and X. H. Chen, Nat. Commun. \textbf{2}, 507 (2011).
\bibitem{Kx1289} M. Xue, T. Cao, D. Wang, Y. Wu, H. Yang, X. Dong, J. He, F. Li, and G. F. Chen, Sci. Rep. \textbf{2}, 389 (2012).
\bibitem{chenxiaojia120K} R. S. Wang, Y. Gao, Z. B. Huang and X. J. Chen, arXiv:1703.06641.
\bibitem{Dessau} H. X. Li,	X. Q. Zhou,	S. Parham, T. Nummy, J. Griffith, K. Gordon, E. L. Chronister and D. S. Dessau, arXiv:1704.04230.
\bibitem{FengDL} M. Q. Ren, W. Chen, Q. Liu, C. Chen, Y. J. Qiao, Y. J. Chen, G. Zhou, T. Zhang, Y. J. Yan and D. L. Feng, arXiv:1705.09901.
\bibitem{LiOHFeSe} H. Lin, J. Xing, X. Y. Zhu, H. Yang, and H. H. Wen, Sci. China-Phys. Mech. Astron. \textbf{59}, 657404 (2016).

\end{thebibliography}
\end{document}